\documentstyle[aps,eqsecnum]{revtex}
\input epsf
\def\tR{{\tilde R}}
\def\Rp{{R'}}
\def\tF{{\tilde F}}
\def\tF{{\tilde F}}
\def\tPsi{{\tilde \Psi}}

\def\dtau{{\dot \tau}}
\def\drho{{\dot \rho}}
\def\taup{{\tau'}}
\def\rhop{{\rho'}}
\def\bP{{\overline P}}
\def\cF{{\cal F}}

\def\cG{{\cal G}}

\def\cW{{\cal W}}
\def\cU{{\cal U}}

\def\hP{{\hat P}}
\def\cOi{{\cal O}^\infty}
\def\cO{{\cal O}}
\def\oP{{\overline P}}

\begin{document}
\twocolumn[\hsize\textwidth\columnwidth\hsize\csname
@twocolumnfalse\endcsname
\title{Toward a Midisuperspace Quantization of LeMa\^\i tre-Tolman-Bondi Collapse Models\\}
\author{Cenalo Vaz${}^{a}$ and Louis Witten$^{b}$ and T.P. Singh${}^{c}$ \\}

\address{$^{a}$Unidade de Ci\^encias Exactas, \\
Universidade do Algarve, Faro, Portugal.\\
{\rm Email address: cvaz@ualg.pt}}

\address{$^{b}$Department of Physics,\\
University of Cincinnati, Cincinnati, OH 45221-0011, USA.\\
{\rm Email address: witten@physics.uc.edu}}

\address{$^{c}$Tata Institute of Fundamental Research,\\
Homi Bhabha Road, Mumbai 400 005, India.\\
{\rm Email address: tpsingh@nagaum.tifr.res.in}}

\maketitle
\thispagestyle{empty}
\begin{abstract}
LeMa\^\i tre-Tolman-Bondi models of spherical dust collapse have been used and
continue to be used extensively to study various stellar collapse scenarios.
It is by now well-known that these models lead to the formation of black holes
and naked singularities from regular initial data. The final outcome of the
collapse, particularly in the event of naked singularity formation, depends
very heavily on quantum effects during the final stages. These quantum effects
cannot generally be treated semi-classically as quantum fluctuations of the
gravitational field are expected to dominate before the final state is
reached. We present a canonical reduction of LeMa\^\i tre-Tolman-Bondi
space-times describing the marginally bound collapse of inhomogeneous dust, in
which the physical radius, $R$, the proper time of the collapsing dust,
$\tau$, and the mass function, $F$, are the canonical coordinates, $R(r)$,
$\tau(r)$ and $F(r)$ on the phase space. Dirac's constraint quantization leads
to a simple functional (Wheeler-DeWitt) equation. The equation is solved and the
solution can be employed to study some of the effects of quantum gravity during
gravitational collapse with different initial conditions.
\end{abstract}

\vskip2pc]

\baselineskip10.8pt

\section{Introduction}
In the standard treatment of Hawking radiation\cite{hawk1} from black holes, one begins
with a black hole that is formed in some classical model of gravitational
collapse, surrounds the hole by a quantum field and examines the
propagation of this field on the classical background provided by the hole.
The quantum field behaves as a thermometer. The quantum modes within the
horizon are averaged over and one finds that the black hole radiates
thermally leading to its ``evaporation''. The temperature that characterizes
the evaporation of a black hole is inversely proportional to its mass and
the radiation flux is inversely proportional to its mass square. For an
astrophysical black hole, the radiation flux and temperature are therefore
so small that the semi-classical approximation is expected to be an adequate
description of the evaporation until the hole is roughly of Planck dimensions at
which point higher order quantum gravity effects will undoubtedly become important.
This means that the final state of the black hole will depend on quantum
gravity. The black hole may evaporate completely or it may leave a remnant. If
it evaporates completely, it is important to understand what happens to the
information that was initially trapped within its horizon.

Classical models of collapse also lead to the formation of naked singularities
for regular initial data\cite{nsform}. One may then ask if the quantum radiation from
naked singularities is similar to the radiation from black holes. By surrounding
a classical naked singularity with a quantum field and examining its quantum modes at null
infinity one finds that the evaporation of a naked singularity is qualitatively distinct
from that of a black hole\cite{FoPa1,HiEa1,TpCv1,TpCv2}. The radiation flux diverges as the Cauchy
horizon\cite{SbTpCvLw1,SbTpCvLw2,SbTpCv1,Har1,Har2} is approached
and the spectrum of the radiation is non-thermal\cite{TpCv1,CvLw1}, falling off as the inverse
frequency. Contrary to the case of an astrophysical black hole, the (divergent)
flux should be observable and the unique spectrum should serve to distinguish objects
undergoing this type of collapse from other celestial emitters. However, a closer look
at the semi-classical approximation just described reveals that the flux of radiation is
essentially negligible (on the order of one Planck mass) until about one Planck
time before the putative Cauchy horizon is reached\cite{Thetal}. The semi-classical
approximation therefore signals a quantum instability of naked singularities (and therefore
a mechanism for the Cosmic Censor) but its quantitative predictions must be tested
in a full quantization of {\it all} the degrees of freedom, including the gravitational
field. When quantum gravitational effects are accounted for, does the flux continue
to diverge and the radiation spectrum continue unique? Indeed, does the quantum
theory serve as a Cosmic Censor and are there significant observational consequences
of collapse into naked singularities as the semi-classical approximation suggests?

In a step toward answering these and other questions regarding the final stages of
gravitational collapse Kucha\v r\cite{ku1} examined a midi-superspace quantization of the
Schwarzschild black hole, presenting in the process a remarkable series of canonical
transformations that greatly simplified the dynamical equations. In the present paper, we
describe a generalization of this work to spherically symmetric, marginally bound
LeMa\^\i tre-Tolman-Bondi space-times. We show below that there is an analogous description
of the gravitational part of the action in terms of the ``mass function'', the physical
radius and their conjugate momenta. Furthermore, the coupling to dust introduces the
proper time of the collapsing matter and its conjugate momentum. Thus time evolution
appears naturally into the dynamical constraints.

The hypersurface action yields two constraints, {\it viz}., the Hamiltonian constraint and the
momentum constraint, which are given in terms of this canonical chart consisting of the mass,
$F[\rho(r)]$, contained within spherical shells of fixed shell-label coordinate, $\rho(r)$,
the physical radius, $R(r)$, the dust proper time, $\tau(r)$, and their conjugate momenta,
$P_F(r)$, $P_R(r)$ and $ P_\tau(r)$ respectively. Here $r$ is the radial label coordinate of
a foliation of the space-time by spacelike hypersurfaces. The momentum conjugate to the mass function,
$P_F(r)$, may be eliminated in the Hamiltonian constraint using the momentum constraint. This
leads to a new and simpler constraint that is able to take the place of the original Hamiltonian
constraint. Dirac's constraint quantization then yields a two dimensional Klein-Gordon-like
functional equation with a potential term that depends on the mass function and its derivative
with respect to the label coordinate $r$. The simplest possible scenario, in which the
mass function is constant throughout the space-time, the same for all shells, describes the
Schwarzschild black hole. The Schwarzschild black hole will thus emerge as a special case of
the general class of models we quantize below.

In section II we review the classical collapse models being considered in this article and
present the canonical formulation for spherically symmetric space-times in section III,
where we also discuss the fall-off conditions appropriate to the models and the resulting
boundary terms. We reconstruct the mass and time from the canonical data in section IV. This
leads naturally to new variables {\it viz.,} the mass, the dust proper time, the physical radius
and their conjugate momenta which are introduced along with the generator of the canonical
transformation from the old to the new variables. We apply Dirac's quantization program to the
new constraints in section V, showing how the new variables lead to a simplified
Wheeler-DeWitt equation. We then present a solution for an arbitrary, but non-constant
mass function.

\section{Classical Models}

The LeMa\^\i tre-Tolman-Bondi models\cite{LTB} constitute a complete solution of the Einstein equations
for a matter continuum of inhomogeneous dust, {\it i.e.,} they are solutions of the spherically
symmetric Einstein's field equations, $G_{\mu\nu} = - 8\pi G T_{\mu\nu}$, with vanishing
cosmological constant and with stress-energy describing inhomogeneous, pressureless dust given
by $T_{\mu\nu} = \epsilon u_\mu u_\nu$. The solution is the LTB metric, given in co-moving
coordinates as\cite{Pj1}
\begin{eqnarray}
ds^2~~ &=&~~ d \tau^2~ -~ {{\tR^2} \over {1 + f}} d\rho^2~ -~ R^2 d\Omega^2,\cr
\epsilon~~ &=&~~ {{\tF}\over {R^2 \tR}},~~ R^*~~ =~~ \pm \sqrt{f~ +~ {F \over R}},
\label{ltbsol}
\end{eqnarray}
(we have set $8\pi G = 1 = c$) where $R$ is the physical radius. A tilde ($~{\tilde {}}~$) represents
a derivative with respect to $\rho$ and a star (${}^*$) represents a derivative with respect
to the dust proper time $\tau$. The functions $f(\rho)$ and $F(\rho)$ are arbitrary functions
only of $\rho$, interpreted respectively as the energy and mass functions. The energy density
of the collapsing matter is $\epsilon(\tau,\rho)$, and the negative sign in the third equation above is
required to describe a collapsing cloud. Its general solution is given up to an arbitrary function
$\psi(\rho)$ of the shell label coordinate. This arbitrariness reflects only a freedom in our choice
of units {\it i.e.,} at any given time, say $\tau_o$, the function $R(\tau_o,\rho)$ can be chosen
to be an arbitrary function of $\rho$.

The mass function, $F(\rho)$, represents the weighted mass
(weighted by the factor $\sqrt{1+f}$) contained within the matter
shell labeled by $\rho$. If a scaling is chosen so that the
physical radius coincides with the shell label coordinate,
$\rho$, at $\tau=0$, then it can be expressed in terms of the
energy density at $\tau=0$ according to
\begin{equation}
F(\rho)~~ =~~ \int \epsilon(0,\rho) \rho^2 d\rho,
\end{equation}
while the energy function, $f(\rho)$, can be expressed in terms of the initial velocity profile,
$v(\rho) = R^*(0,\rho)$, according to
\begin{equation}
f(\rho)~~ =~~ v^2(\rho)~ -~ {1 \over \rho} \int \epsilon(0,\rho) \rho^2 d\rho.
\end{equation}
The marginally bound models, which we will consider in this paper, are defined by $f(\rho)=0$.
For the scaling referred to above, we must choose $\psi(\rho) = \rho^{3 \over 2}$, whence the
solution of (\ref{ltbsol}) can be written as
\begin{equation}
R^{3 \over 2}(\tau,\rho)~~ =~~ \rho^{3 \over 2}~ -~ {3 \over 2} \sqrt{F(\rho)}\tau.
\label{ltbrad}
\end{equation}
The epoch $R=0$ describes a physical singularity, whose singularity curve
\begin{equation}
\tau(\rho)~~ =~~ {{2\rho^{3 \over 2}} \over {3\sqrt{F(\rho)}}},
\end{equation}
gives the proper time when successive shells meet the central physical singularity.
Various models are obtained from choices of the mass function, $F(\rho)$. For example, the
Schwarzschild black hole is the marginally bound solution with $F(\rho) = 2M$, a constant.

A collapsing star does not have sharp boundaries. In the simplest possible approximation
to reality, it will consist of a dense core surrounded by a crust of lower density which,
itself, is encased in a cloud whose density will smoothly go to zero with distance from the
star's center\cite{Pjetal}. Nevertheless, as an approximation, one could consider a sharp
boundary at some constant shell label, $\rho_b$, with an exterior Schwarzschild metric.
One could also consider several regions described by different mass functions describing
successively lower matter densities as one moves out from the center. The mass function
as a whole would not be differentiable at the boundary or boundaries, but we require it to
be continuous.

One can then examine outgoing families of non-spacelike geodesics and check if there exist
congruences that terminate in the past at the central singularity\cite{TpPj1}. If such
congruences exist then the collapse leads to a naked singularity and if they do not exist then
the collapse leads to a black hole. A naked singularity may further be characterized as
globally naked or locally naked depending on whether the outgoing geodesics succeed in
reaching null infinity or not.

In general one finds that both black holes and naked singularities may develop as the end
states of collapse, depending on initial data, {\it i.e.,} on the initial density and velocity
profiles of the collapsing dust. For example, in the marginally bound self-similar collapse
model, $F(\rho) = \lambda \rho$, where $\lambda = {\rm const.}$, both outcomes described by
the Penrose diagrams in figure 1 are possible, depending on whether $\lambda > 0.1809$ (black
hole) or $\lambda \leq 0.1809$ (naked singularity).
\vskip 5mm
\centerline{\epsfysize=4.5cm \epsfbox{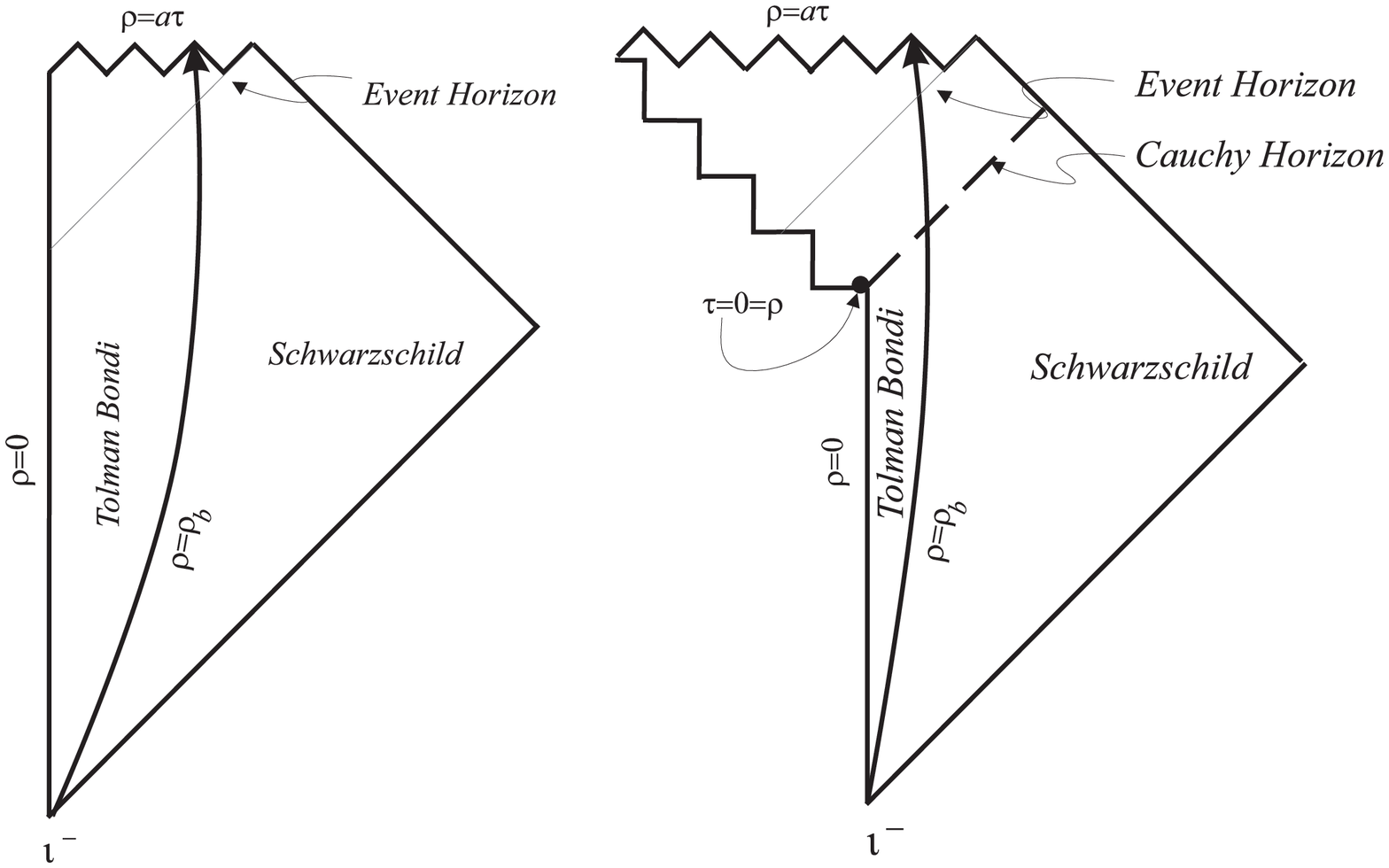}}
\vskip 1mm
\centerline{\small {\bf Fig.1:} Black hole (left) or naked singularity (right)}
\vskip 5mm
\noindent It is in fact believed by many that sets of initial data may evolve in general
relativity toward either naked singularities or black holes independently of the equation of
state or the type of matter used.

Geometrodynamics views space-time as the dynamical evolution of spatial hypersurfaces. When
the collapse evolves toward a black hole such spatial hypersurfaces exist, starting
at infinity, crossing the horizon and continuing to $\rho=0$ without encountering
the central singularity. On the contrary, when the collapse evolves toward a naked singularity,
the spatial hypersurfaces in the future of the initial singularity cross the Cauchy horizon and
collide with the central singularity. No sensible boundary conditions can be specified
on a singularity and evolution in the future of the initial singularity is arbitrary. To avoid
the consequent breakdown of predictability, Penrose proposed the Cosmic Censor\cite{Rp1} which, as
mentioned in the introduction, is likely to be the quantum theory itself and will come into
play {\it before} the Cauchy horizon has a chance to form. It is of particular interest,
therefore, to understand precisely how the system behaves close to, but in the past of, the
putative Cauchy horizon. Spatial hypersurfaces in the past of the Cauchy horizon are well
defined and the quantum evolution of the system may be studied until the time of
its formation.

\section{Canonical Dynamics}

The line element $d\sigma$ on a spherically symmetric three-dimensional Riemann surface
$\Sigma$ is completely characterized by two functions, $L(r)$ and $R(r)$ of the radial
label coordinate, $r$, according to
\begin{equation}
d\sigma^2~~ =~~ L^2(r) dr^2~ +~ R^2(r) d\Omega^2
\end{equation}
where $\Omega$ is the solid angle. Neither $L(r)$ nor $R(r)$ can be negative and we take them
to be positive definite except possibly at the center. $R(r)$ represents the physical radius
of the point labeled by $r$ on the surface. It behaves as a scalar under transformations of
$r$, whereas $L(r)$ behaves as a scalar density. The corresponding four dimensional line element
may be written in terms of two additional functions, the lapse, $N(t,r)$, and the shift,
$N^r(t,r)$, as
\begin{equation}
ds^2~ =~ N^2 dt^2 - L^2(dr - N^r dt)^2 - R^2 d\Omega^2.
\label{gsmet}
\end{equation}
In this spherically symmetric space-time, we will consider the Einstein-Dust system
described by the action
\begin{eqnarray}
S = & - & {1 \over {16\pi}} \int d^4 x \sqrt{-g}~ {\cal R}\cr
    & - & {1 \over {8\pi}} \int d^4 x \sqrt{-g}~ \epsilon(x) \left[g_{\alpha\beta}
        U^\alpha U^\beta + 1 \right]
\label{action}
\end{eqnarray}
where ${\cal R}$ is the scalar curvature. As is well known, the gravitational part of this
action can be cast into the form
\begin{eqnarray}
S^g =  \int  &dt& \int_0^\infty dr \left [ P_L {\dot L} + P_R {\dot R} - N H^g
- N^r H^g_r \right]\cr\cr
& + &~ S^g_{\partial \Sigma}
\end{eqnarray}
with the momenta conjugate to $L$ and $R$ respectively given by
\begin{eqnarray}
P_L~~ &=&~~ {R \over N} \left[-{\dot R} + N^r R'\right]\cr
P_R~~ &=&~~ {1 \over N} \left[-L{\dot R} - {\dot L} R + (N^r LR)'\right]
\label{gravmom}
\end{eqnarray}
and where the overdot and the prime refer respectively to partial derivatives with respect
to the label time, $t$, and coordinate, $r$. The lapse, shift and phase-space variables are
required to be continuous functions of the label coordinates. The boundary action,
$S^g_{\partial \Sigma}$, is required to cancel unwanted boundary terms in the hypersurface
action, thereby ensuring that the hypersurface evolution is not frozen. It is determined
after fall-off conditions appropriate to the models under consideration are specified. The
super-Hamiltonian and super-momentum constraints are given by
\begin{eqnarray}
H^g = &-& \left[{{P_L P_R} \over R} - {{LP_L^2} \over {2R^2}}\right]\cr
&+& \left[ - {L \over 2} - {{{R'}^2} \over {2L}} + \left({{RR'} \over L}\right)'\right]\cr\cr
H^g_r = &R'& P_R - L P_L'
\label{const}
\end{eqnarray}
We will assume that the chosen mass function, $F(\rho)$, is such that at infinity Kucha\v r's
fall-off conditions\cite{ku1} are suitable and we will adopt them here. These conditions would
be applicable, for example, in models in which the collapsing metric asymptotically approaches
or is smoothly matched to an exterior Schwarzschild background at some boundary, $\rho_b$. They
read
\begin{eqnarray}
L(t,r)~~ &=& ~~ 1~ +~ M_+(t) r^{-1}~ +~ \cOi(r^{-1-\epsilon})\cr
R(t,r)~~ &=& ~~ r~ +~ \cOi(r^{-\epsilon})\cr
P_L(t,r)~~ &=& ~~ \cOi(r^{-\epsilon})\cr
P_R(t,r)~~ &=&~~ \cOi(r^{-1-\epsilon})\cr
N(t,r)~~ &=& ~~ N_+(t)~ +~ \cOi(r^{-\epsilon})\cr
N^r(t,r)~~ &=&~~ \cOi(r^{-\epsilon})
\label{foinf}
\end{eqnarray}
Again, because the label radial coordinate $r \in [0,\infty)$, we must also consider the boundary
conditions at $r=0$. Let the mass function near the center ($\rho=0$) have a series expansion of
the form
\begin{equation}
F(\rho)~~ =~~ F_o + F_1 \rho + F_2 \rho^2 + ... = \sum_n F_n \rho^n.
\end{equation}
If $F_n = 0~ \forall~ n > 0$ but $F_o > 0$, the solution describes the Schwarzschild black
hole of mass $F_o/2$. The marginally bound, self-similar model mentioned in the previous section
corresponds to $F_1 = \lambda > 0$ and $F_n = 0~ \forall~ n \neq 1$, or a density profile
that behaves as $\epsilon(0,\rho) \sim \rho^{-2}$. Thus $\tau=0$ is the singular epoch for the
self-similar model and this singular density profile arises from a regular initial profile at
some $\tau < 0$.

Consider models with $F(\rho) = F_n \rho^n$. As far as the fall-off conditions at the center are
concerned, two classes of models arise, {\it viz.,} $n \leq 3$ and $n>3$. The
considerations below are applicable to models with $n\leq3$, although conditions appropriate to
models with $n > 3$ may likewise be given. Referring to equation (\ref{ltbrad}) we find that as
$\rho \rightarrow 0$, $R(\tau,\rho) \approx R_o(\tau) \rho^{n\over 3} + \cO(\rho^{{n \over 3} + k})$
and $L(\tau,\rho) \approx L_o(\tau)\rho^{{n\over 3}-1} + \cO(\rho^{{n \over 3} -1 + k})$, where $k>0$.
We will exclude the black hole ($n=0$) in the following because in that case the space-time may be
analytically continued to $r=-\infty$ where the boundary conditions given in (\ref{foinf}) may be
applied and no conditions at $r=0$ need be given. For a genuine collapse, the cases of interest
are those with $n>0$ around the central region.

Let us assume that, at the center, $(\rho,\tau)$ approach $(r,t)$ as $\rho = r + \cO(r^{\alpha})$
($\alpha > 1$) and $\tau = f(t) + \cO(r^{\alpha})$. Now, in the following section we will show that
the mass function is recovered locally from the canonical data according to
\begin{equation}
F=R\left[1 + {{P_L^2} \over {R^2}} - {{\Rp^2} \over {L^2}}\right].
\end{equation}
Again, with $\rho = r + \cO(r^{\alpha})$ and $\tau = f(t) + \cO(r^{\alpha})$, it follows that
\begin{equation}
{\Rp \over L} \approx 1 + \cO(r^{\alpha -1}).
\end{equation}
and therefore
\begin{eqnarray}
F = F_n \rho^n &=& F_n r^n + \cO(r^{n+\alpha -1}) \cr
    &=& {{P_L^2} \over R} + \cO(r^{n/3 + \alpha -1}),
\end{eqnarray}
or
\begin{equation}
F_n = {{P_L^2} \over {R_o}} r^{-4n/3} + \cO(r^{-2n/3 + \alpha -1}).
\end{equation}
The last equation can be satisfied if $P_L$ falls off faster than $r^{2n/3}$ and $\alpha = 1 + 2n/3$.
Furthermore, requiring both terms in the Liouville form to have the same behavior at the origin,
we choose the following conditions near $r=0$ when $n=\{1,2,3\}$:
\begin{eqnarray}
R(r,t) &=& R_o(t) r^{n/3} + \cO(r^{n/3+\epsilon})\cr
L(r,t) &=& L_o(t) r^{n/3-1} + \cO(r^{n/3-1+\epsilon})\cr
P_L(r,t) &=& \cO(r^{2n/3 + 1 + \epsilon})\cr
P_R(r,t) &=& \cO(r^{2n/3 + \epsilon})\cr
N(r,t)~~ &=& N_o(t) + \cO(r^{\epsilon})\cr
N^r(r,t) &=& \cO(r^{\epsilon})
\label{fo0}
\end{eqnarray}
The conditions (\ref{foinf}) and (\ref{fo0}) ensure that the Liouville form is well behaved both
at the origin ($\cO(r^{(n+\epsilon)})$) as well as at infinity ($\cOi(r^{-\epsilon})$). The
Hamiltonian and momentum densities behave as
\begin{equation}
H^g = \cO(r^{n+1+\epsilon}),~~ H^g_r = \cO(r^{n-1+\epsilon})
\label{fo0H}
\end{equation}
as the origin is approached, and
\begin{equation}
H^g = \cOi(r^{-2(1+\epsilon)}),~~ H^g_r = \cOi(r^{-(1+\epsilon)})
\label{foinfH}
\end{equation}
asymptotically. Thus the total Hamiltonian and momentum are well defined and the surface action is
meaningful. The potential contributions to the surface action can be read off the constraint
equations, (\ref{const}), by considering variations of the phase space variables. Applying the
fall-off conditions at infinity, one finds that only one of these is non-vanishing and behaves as
\begin{equation}
\int_{\partial\Sigma_\infty} dt N_+(t) \delta M_+(t).
\end{equation}
On the other hand, with the fall-off conditions in (\ref{fo0}) as $r\rightarrow 0$, all the
variations vanish at the origin. Therefore the only contribution to the boundary action comes
from the surface term at infinity and we find
\begin{equation}
S_{\partial \Sigma}~~ =~~ -\int_{\partial \Sigma_\infty} dt N_+(t) M_+(t).
\end{equation}
We will return to this surface action shortly.

Let us now turn to the dust portion of the action in(\ref{action}),
\begin{equation}
S^d~~ =~~ \int d^4 x \sqrt{-g}~ \epsilon(x)\left[g_{\alpha\beta} U^\alpha U^\beta + 1
\right].
\end{equation}
It has been exhaustively analyzed by Kucha\v r and Torre\cite{KuCt} and by Brown and
Kucha\v r\cite{JbKu}. It may be understood in two ways: either as a consequence of imposing
coordinate conditions (the Gaussian conditions) or as a realistic material medium. For
the collapse problem it is a realistic material medium and for the LTB models being considered,
it is non-rotating. Dust is described by eight space-time scalars, $\epsilon$, $\tau$, $Z^k$ and
$W_k$ ($k~ \in~ \{1,2,3\}$). The physical interpretation of these variables which follows from
an analysis of the equations of motion was given in ref. \cite{JbKu} and will be summarized here for
completeness. $\tau$ is the proper time measured along particle flow lines, $Z^k$ are the comoving
coordinates of the dust, $W^k$ are the spatial components of the four velocity in the dust frame,
and $\epsilon$ is the dust proper energy density. All these scalars are assumed to be functions
of the space-time coordinates. In particular, the four variables, $Z^K = (\tau, Z^k)$, are independent
functions, ${\rm det}|Z^K_{~~,\mu}| \neq 0$, and the four-velocity of the dust particles may be
defined by its decomposition in the co-basis
$Z^K_{~~,\mu}$ by
\begin{equation}
U_\mu~~ =~~ -\tau_{,\mu}~ +~ W_k Z^k_{~~, \mu}.
\end{equation}
In the spherically symmetric geometry described by (1.5), the dust action may be
cast into the form
\begin{equation}
S^d~~ =~~ \int dt \int dr  \left[P_\tau {\dot \tau} + P_k {\dot Z}^k - N H^d - N^r H^d_r
\right],
\end{equation}
where
\begin{eqnarray}
P_\tau~~ &=&~~ {{LR^2} \over N} \epsilon(r,t) \left[-U_t + N^r U_r\right]\cr
P_k~~ &=&~~  - W_k P_\tau
\label{dustmom}
\end{eqnarray}
are the momenta conjugate to the dust proper time and the frame
variables respectively, and
\begin{eqnarray}
H^d~~ &=&~~ P_\tau \sqrt{1 + {{U_r^2} \over {L^2}}}\cr
H^d_r~~ &=&~~ - U_r P_\tau.
\label{dustconst}
\end{eqnarray}
This expression for $H^d$ is obtained by exploiting the fact that $\epsilon(t,r)$ is a Lagrange
multiplier, and therefore $\delta {\cal L} / \delta \epsilon = 0$. When the dust is non-rotating
({\ref{dustconst}) is further simplified by requiring that the dust motion be described with respect
to the frame orthogonal foliation. Then we may impose the additional constraints $P_k=0$ (when
imposed on the state functional these constraints simply mean that the state functional does not
depend on the frame variables $Z^k$). Using this in (\ref{dustmom}) we see that $W_k = 0$ and so
$U_r = -\taup$.

Combining the results thus far obtained, we now give the full form of the canonical description of
the action in (\ref{action}) as:
\begin{eqnarray}
S = \int  dt \int_0^\infty dr~ [&P_\tau& \dtau + P_L {\dot L}\cr
&+& P_R {\dot R} - N H - N^r H_r ] + S^g_{\partial \Sigma}
\end{eqnarray}
with
\begin{eqnarray}
H = &-& \left[-{{P_L P_R} \over R} - {{LP_L^2} \over {2R^2}}\right]\cr
&+& \left[ - {L \over 2} - {{{R'}^2} \over {2L}} + \left({{RR'} \over L}\right)'\right]\cr
&+& P_\tau \sqrt{1 + {{\taup^2} \over {L^2}}}~~ \approx~~ 0\cr\cr
H_r = &\tau'& P_\tau + R' P_R - L P_L'~~ \approx~~ 0
\end{eqnarray}
and
\begin{equation}
S_{\partial \Sigma}~~ =~~ -\int_{\partial \Sigma_\infty} dt N_+(t) M_+(t).
\end{equation}
for the boundary contribution.

\section{New Variables}

The hypersurfaces we consider, from which (\ref{gsmet}) is constructed, must eventually
be embedded in a space-time described by the metric given in (\ref{ltbsol}) with $f=0$. We
imagine that they are leaves of the foliation $\tau(t,r)$ and $\rho(t,r)$. Then the
functions $L(t,r)$ and $R(t,r)$ appearing in (\ref{gsmet})are easily determined by
substituting the foliation in (\ref{ltbsol}). We find
\begin{eqnarray}
L^2~~ &=&~~ \tR^2 \rhop^2~ -~ \taup^2\cr
N^r~~  &=&~~ {{\tR^2 \drho \rhop - \dtau \taup} \over L^2}\cr
N~~ &=&~~ {\tR \over L} (\dtau \rhop~ -~ \drho \taup)
\label{embed}
\end{eqnarray}
The last of the equations above involves taking a square root. We must check that
the positive square root taken leads to a positive lapse function in all the regions
of the space-time. Call $\cF = 1 - F/R$ where $F$ is the mass function and note that
$R^* = -\sqrt{1-\cF}$ according to (\ref{ltbsol}) with $f=0$.

Substituting the expression for $N(t,r)$ and $N^r(t,r)$ obtained in (\ref{embed})
into the expressions for the canonical momenta in (\ref{gravmom}), one obtains
the relation
\begin{eqnarray}
{{LP_L} \over R}~ =~ {1 \over {\tR ({\dot \tau} \rho'-\tau'{\dot \rho})}}
[ &-& {\dot R} (\tR^2 \rhop^2 - \taup^2) \cr
&+& ~ R' (\tR^2 \drho \rhop- \dtau\taup)],
\end{eqnarray}
which, after some algebra, can be used to obtain $\taup$ in terms
of the canonical variables,
\begin{equation}
\tau'~~ =~~ -~ {{LP_L} \over {R\cF}}~ +~ {{R'\sqrt{1-\cF}} \over \cF}.
\label{ptimefn}
\end{equation}
Inserting this into the expression for $L^2$ in (\ref{embed}), we find that
\begin{equation}
L^2~~ =~~ {{\Rp^2} \over \cF}~ -~ {{L^2 P_L^2} \over {R^2 \cF}},
\end{equation}
which determines $\cF$,
\begin{equation}
\cF~~ =~~ {{\Rp^2} \over {L^2}}~ -~ {{P_L^2} \over {R^2}},
\end{equation}
and, through $\cF$, recovers the mass function
\begin{equation}
F~~ =~~ R\left[1~ +~ {{P_L^2} \over {R^2}}~ -~ {{\Rp^2} \over {L^2}} \right].
\end{equation}
This is the relation used in the previous section when discussing the fall-off conditions at
the center. It enables us to determine the mass function {\it locally} from the canonical data.
Furthermore, once the dust proper time is fixed at some point on the hypersurface, say at
spatial infinity, equation (\ref{ptimefn}), which determines the difference in dust proper
times between any two points $r_1$ and $r_2$ on a spatial hypersurface, will determine it at
any point on the hypersurface. Note that at the horizon, when $\cF=0$,
{\it i.e.,}
\begin{equation}
{\Rp \over L} =  {{P_L} \over R},
\end{equation}
$\taup$ continues to be well behaved, as expected.

It turns out that the functions $P_F$, defined by\footnote{$P_F(r)$ is the equivalent of
Kucha\v r's $P_M(r)$,the momentum conjugate to the black hole mass function, $M(r)$. The
parallel between our construction for LTB metrics and Kucha\v r's construction for the
Schwarzschild black hole, which inspired this work, is remarkable because the Schwarzschild
metric (with a varying mass, $M(r)$) is not diffeomorphic to the LTB metric in (\ref{ltbsol})
{\it except} in the black hole case, {\it i.e.,} when
$F=2M={\rm const.}$, and $f=0$.}
\begin{equation}
P_F = {{LP_L} \over {2R\cF}},
\end{equation}
and the mass function, $F$, form a conjugate pair of variables. Moreover, because neither $F$
nor $P_F$ depend on $P_R$, they have vanishing Poisson brackets with $R$. They also have
vanishing Poisson brackets with $\tau$ and $P_\tau$. Their Poisson brackets with $P_R$, however,
do not vanish and one cannot directly replace the pair $(L,P_L)$ with the more transparent
variables $(F,P_F)$ to form a new chart. Instead one asks if it is possible to determine a new
momentum, $\oP_R$, conjugate to $R$ and such that the set $(\tau,R,F,P_\tau,\oP_R,P_F)$ forms a
canonical chart. We proceed by constructing $\oP_R$ in exactly the same way as Kucha\v r did for
the Schwarzschild black hole. Then we show that the transition to the new chart is indeed a
canonical transformation by displaying its generator.

Kucha\v r\cite{ku1} proposed that $(R,F,\oP_R,P_F)$ should form a canonical chart whose coordinates
are spatial scalars, whose momenta are scalar densities and which is such that $H_r(r)$
generates Diff {\bf R}. This means that
\begin{eqnarray}
H_r &=& \tau'P_\tau + R'P_R - L P_L'\cr\cr
&=& \tau'P_\tau + R'\oP_R + F'P_F \approx 0
\end{eqnarray}
Substituting the expressions derived earlier for $F$ and $P_F$ into the above one finds
\begin{equation}
\oP_R = P_R - {{LP_L} \over {2R}} - {{LP_L} \over {2R\cF}} - {{\Delta}
\over {RL^2 \cF}}
\label{op}
\end{equation}
where $\Delta = (RR')(LP_L)' - (RR')'(LP_L)$. We must now show that the transformation
\begin{equation}
(\tau,R,L,P_\tau,P_R,P_L) \rightarrow (\tau,R,F, P_\tau,\oP_R,P_F)
\end{equation}
is a canonical transformation.

To do this we solve the set
\begin{eqnarray}
p_i(r) &=& \sum_j \int_0^\infty dr' P_j(r') {{\delta Q_j(r')}
\over {\delta q_i(r)}} + {{\delta \cG} \over {\delta q_i(r)}}\cr 0
&=& \sum_j \int_0^\infty dr' P_j(r') {{\delta Q_j(r')} \over
{\delta p_i(r)}} + {{\delta \cG} \over {\delta p_i(r)}}
\end{eqnarray}
where $(p_i,q_i)$ and $(P_i,Q_i)$ are respectively the old and the new phase-space variables
and where $\cG[q_i,p_i]$ generates the transformation. Because we know $F$ and $P_F$ in terms
of $(R,L,P_R,P_L)$, the four non-trivial equations, {\it viz.,}
\begin{eqnarray}
P_L(r) &=& \int_0^\infty dr' P_F(r') {{\delta F(r')} \over {\delta
L(r)}} + {{\delta \cG} \over {\delta L(r)}}\cr\cr P_R(r) &=&
\oP_R(r) + \int_0^\infty dr' P_F(r'){{\delta F(r')} \over {\delta
R(r)}} + {{\delta \cG} \over {\delta R(r)}}\cr\cr 0 &=&
\int_0^\infty dr' P_F(r') {{\delta F(r')} \over {\delta P_L(r)}} +
{{\delta \cG} \over {\delta P_L(r)}}\cr\cr 0 &=& \int_0^\infty dr'
P_F(r') {{\delta F(r')} \over {\delta P_R(r)}} + {{\delta \cG}
\over {\delta P_R(r)}}, \label{cantrans}
\end{eqnarray}
can be solved for $\cG$, and $\oP_R$ can be recovered using the second of the above equations.
The last of (\ref{cantrans}) implies that $\cG$ is independent of $P_R$. The third equation in
(\ref{cantrans}) reads
\begin{equation}
{{\delta \cG} \over {\delta P_L}} = - {{2P_F P_L} \over R} = - {{L P_L^2} \over {R^2}}
\left[{{\Rp^2} \over {L^2}}~ -~ {{P_L^2} \over {R^2}}\right]^{-1}.
\end{equation}
and can be integrated to give
\begin{eqnarray}
\cG = \int_0^\infty &dr& \left[LP_L - {1 \over 2}{RR'} \ln |{{RR'+LP_L} \over {RR'-LP_L}}|
\right]\cr\cr
&+& \cG_2[R,L]
\label{genfn}
\end{eqnarray}
and the first equation ensures that $\cG_2$ is independent of $L$. We will take $\cG_2$ to be
independent also of $R$ and use $\cG$ to determine $\oP_R$ from the second equation in
(\ref{cantrans}). This gives precisely (\ref{op}). Next, we verify that the transformation
has not introduced fresh boundary terms by computing the difference between the
old and the new Liouville forms,
\begin{eqnarray}
\int_0^\infty dr [&P_R& \delta R + P_L \delta L - \oP_R \delta R - P_F \delta F] = \cr\cr
\int_0^\infty &dr& \left\{\delta\left[LP_L - {1 \over 2}{RR'} \ln |{{RR'+LP_L} \over {RR'-LP_L}}|
\right]\right. \cr\cr
&&~~ \left. + \left[ {1 \over 2} R\delta R \ln |{{RR'+LP_L} \over {RR'-LP_L}}|\right]'\right\}.
\end{eqnarray}
Kucha\v r\cite{ku1} has shown that the fall-off conditions (\ref{foinf}) at infinity imply that
\begin{eqnarray}
\left[{1 \over 2} R\delta R \ln |{{RR'+LP_L} \over {RR'-LP_L}}|\right]_{r\approx \infty}
&\approx&  {{LP_L\delta R} \over \Rp}\cr\cr
&\approx& \cO(r^{-\epsilon})
\end{eqnarray}
and hence vanishes at infinity. Again, as $r \rightarrow 0$, the fall-off conditions in (\ref{fo0})
imply that $R \rightarrow R_o(t) r^{n/3}$, $\delta R \rightarrow \delta R_o(t) r^{n/3}$, $R'
\rightarrow R_o(t) r^{n/3-1}$, $L \rightarrow L_o(t) r^{n/3 -1}$ and $P_L \rightarrow \cO(r^{2n/3+
1+\epsilon})$. We find that
\begin{eqnarray}
\left[ {1 \over 2} R \delta R \ln |{{RR'+LP_L} \over {RR'-LP_L}}|\right]_{r\approx 0}
&\approx&  {{LP_L\delta R} \over \Rp}\cr\cr
&\approx& \cO(r^{n+1+\epsilon})
\end{eqnarray}
and therefore also vanishes at the origin. The functional $\cG$ defined by (\ref{genfn}) is
well defined. The integrand is of order $r^{-(1+\epsilon)}$ at infinity and of order
$r^{4n/3+1+\epsilon}$ at the origin, avoiding divergences at both places.

There are (infinite) boundary terms at the horizon, when $\cF = 0$. It can be shown,
however, that the contribution from the interior and the exterior cancel each other. In the
same way, there will be contributions at the boundary between the interior of the star and
its exterior or more generally at any frontier between two LTB regions described by different
mass functions. Again, if the mass function is continuous across the boundary and regions
are consistently matched by equating both the first and second fundamental forms, then the
contribution from one side will cancel the contribution from the other.

In terms of the new variables, the action in (\ref{action}) can be expressed as
\begin{eqnarray}
S = \int dt \int_0^\infty dr [P_\tau {\dot \tau} &+& \oP_R {\dot R} + P_F {\dot F} - \cr\cr
&-& N H^g - N^r H^g_r] + S_{\partial\Sigma}
\end{eqnarray}
with
\begin{eqnarray}
H &=& - \left[{{F' \cF^{-1} R' + \cF P_F \bP_R} \over {2L}}\right] + P_\tau \sqrt{1 +
{{\taup^2} \over {L^2}}}\cr\cr
H_r &=& \taup P_\tau~ + R'\oP_R + F' P_F\cr\cr
S_{\partial\Sigma} &=& -\int_{\partial\Sigma_\infty} N_+(t)M_+(t)
\end{eqnarray}
Let us now re-express the boundary action in a more convenient form. As Kucha\v r \cite{ku1}
has emphasized, $N_+(t)$ must be treated as a {\it prescribed} function of $t$. This is to avoid
the conclusion that the total mass $M_+$ as measured at infinity is zero, which would follow
from varying the lapse function at infinity. $N_+(t)$ must be chosen and, once chosen, held fixed.
The freedom in choosing this function can be combined with the freedom we have of setting the
dust proper time at infinity to correspond to the parameterization clocks there. The lapse
function is the rate of change of the proper time with the
coordinate time at infinity, so it is natural to set $N_+(t) =
{\dot \tau}_+(t)$ and write the surface action as
\begin{equation}
S_{\partial\Sigma} = - \int dt [M_+ {\dot \tau}_+].
\end{equation}
It is linear in the time derivative, ${\dot \tau}_+$, and defines a one form which can be
re-written in terms of the mass function, $F$, as
\begin{eqnarray}
-M_+ \delta \tau_+ &=& - {1\over 2}\int_0^\infty dr (F \delta \tau)'\cr\cr
&=& - {1 \over 2}\int_0^\infty dr [F' \delta \tau - \tau' \delta F + \delta(F\tau')].
\end{eqnarray}
The first two terms on the right hand side may be absorbed into the Liouville form
of the hypersurface action (they modify the canonical momenta) and the last term is
an exact form which can be dropped. The action is thus expressed entirely as
a hypersurface action. Defining $\oP_\tau = P_\tau - F'/2$ and $\oP_F = P_F + \tau'/2$,
we have
\begin{eqnarray}
S = \int dt \int_0^\infty dr [\oP_\tau {\dot \tau} &+& \oP_R {\dot R} + \oP_F {\dot F} - \cr\cr
&-& N H^g - N^r H^g_r]
\end{eqnarray}
where the constraints in the new chart read
\begin{eqnarray}
H &=& - \left[{{F' \cF^{-1} R' + \cF (\oP_F -\tau'/2) \oP_R} \over {2L}}\right]\cr\cr
&&~~ + (\oP_\tau +F'/2) \sqrt{1 + {{\taup^2} \over {L^2}}}\cr\cr
H_r &=& \taup \oP_\tau~ + R'\oP_R + F' \oP_F
\label{finalconstraints}
\end{eqnarray}
Finally, eliminating the momentum $\oP_F$ from the Hamiltonian constraint by using the
momentum constraint, we obtain
\begin{equation}
(\oP_\tau + F'/2)^2 + \cF \oP_R^2 - {{{F'}^2} \over {4\cF}} \approx 0.
\end{equation}
This is the simplified constraint referred to in the introduction. It takes the place of
the Hamiltonian constraint in (\ref{finalconstraints}). In the following section we
quantize the constraints and solve the Wheeler-DeWitt equation for arbitrary non-constant
mass functions.

\section{Quantization}

From the dynamical constraint in the previous section, one reads off the DeWitt supermetric,
$\gamma_{ab}$, on the configuration
space $X^a = (\tau,R)$,
\begin{equation}
\gamma_{ab} = \left(\matrix{1&0\cr0&{1\over \cF}}\right).
\label{confmet}
\end{equation}
To quantize the system, the momenta must be turned into operators which act on the state
functional. There is a standard procedure to do this, which follows a proposal due to
Vilkovisky\cite{Vi1,BaVi1} and DeWitt\cite{Dw1}: exchange the classical momenta for covariant
functional derivatives,
\begin{equation}
\hP_a = - i \nabla_a = -i \left({{\delta} \over {\delta X^a(r)}} + \Gamma_a\right)
\end{equation}
where $\Gamma$ is the connection belonging to the configuration space metric, $\gamma_{ab}$.
Then defining
\begin{equation}
\Psi[\tau,R,F] = e^{-{i\over 2}\int_0^\infty F'(r)\tau(r)  dr} \tPsi[\tau,R,F],
\end{equation}
we see that $\tPsi$ obeys the ``Klein-Gordon'' equation with a potential,
\begin{equation}
\left[\gamma^{ab} \nabla_a \nabla_b + {{{F'}^2} \over {4\cF}}\right] \tPsi[X,F] = 0.
\label{reducedeqn}
\end{equation}
The metric in (\ref{confmet}) is positive definite outside the horizon ($\cF > 0$) and indefinite
inside ($\cF < 0$). The functional equation is therefore elliptic outside the horizon and hyperbolic
inside. Furthermore, the configuration space is flat and this metric is brought to a manifestly
flat form by the coordinate transformation
\begin{equation}
R_* = \int {{dR}\over {\sqrt{|\cF|}}}.
\end{equation}
In terms of $R_*$, the Wheeler-DeWitt equation reads
\begin{equation}
\left[{\delta^2 \over {\delta \tau^2}} \pm {\delta^2 \over {\delta R_*^2}} + {{{F'}^2}
\over {4\cF}} \right] \tPsi[\tau,R, F] = 0,
\label{dyn}
\end{equation}
or
\begin{equation}
i {{\delta \tPsi} \over {\delta \tau}} = {\hat h} \tPsi = \pm \sqrt{\mp{{\hat P}_*}^2 + {{{F'}^2}
\over {4\cF}}}~ \tPsi[\tau,R, F] = 0,
\label{dyn2}
\end{equation}
where the negative sign within the square root refers to the region outside the horizon, the positive
sign to the interior and where ${\hat P}_*$ is conjugate to $R_*$. Invariance under spatial diffeomorphisms
is enforced by the momentum constraint,
\begin{equation}
\left[\tau'{\delta \over {\delta \tau}} + R_*'{\delta \over {\delta R_*}} + F'
{\delta \over {\delta F}}\right] \tPsi[\tau,R,F] = 0,
\label{diff}
\end{equation}
and, together, eqs. (\ref{dyn}) and (\ref{diff}) define the quantum theory whose inner
product is given by
\begin{equation}
\langle \Psi_1, \Psi_2\rangle = \int_{R_*(0)}^\infty dR_* \Psi_1^\dagger
\Psi_2 = \int_{R_*(0)}^\infty \prod_r dR_*(r) \Psi_1^\dagger \Psi_2
\label{inprod}
\end{equation}
where $R(0)$ represents the physical radius at the center, $r=0$. The inner product ensures
the hermiticity of the momentum, ${\hat P}_*$, conjugate to $R_*$. The norm of a quantum state under
this scalar product is formally $\tau$ independent provided that ${\hat h}$ defined in (\ref{dyn2})
is self-adjoint. However, this occurs only in the linear sub-space in which the operator
$\mp{{\hat P}_*}^2 + {F'}^2/ 4\cF$ admits positive eigenvalues. When $F'=0$, for example for the
Schwarzschild black hole, this is true of all states with support only in the interior of the
hole\cite{vw}. Such states alone may be ascribed a probabilistic interpretation.

While specific models of stellar collapse must be analyzed individually, a solution for general $F(r)
\neq$ const. can be given as follows. The momentum constraint requires the wave functional to be a
spatial scalar, therefore consider a solution of the form
\begin{equation}
\tPsi[\tau,R,F] = \exp\left[{1 \over 2} \int_0^\infty dr F'(r)\cW(\tau(r),R(r),F(r))\right]
\end{equation}
where $\cW(\tau,R,F)$ is an arbitrary (complex valued) function of the coordinates (and not their
derivatives) which is to be determined. The integrand in the exponent is clearly a spatial density
because, while $\tau(r)$, $R(r)$ and $F(r)$ are spatial scalars, $F'(r)$ is a density. It follows that
the wave-functional so defined will obey the momentum constraint. Indeed, (\ref{diff}) simply requires
that $\cW$ admit no explicit dependence on $r$. It is well-known that the Wheeler-DeWitt equation requires
some regularization, involving as it does two functional derivatives taken at the same point. We will
follow DeWitt \cite{Dw2} (see also \cite{Hori}) and require that the delta function vanishes in
coincidence limit, as does the coincidence limit of its derivatives to all orders, {\it i.e.,} we let
$\delta(0) = 0 = \delta^{(n)}(0)$ in what follows (where the superscipt $(n)$ refers to the $n^{\rm th}$
spatial derivative of the delta function). Then the dynamical equation (\ref{dyn}) reads
\begin{equation}
{{F'^2} \over 4} \left[\left({{\partial \cW} \over {\partial \tau}}\right)^2 \pm\left(
{{\partial \cW} \over {\partial R_*}}\right)^2 + {1 \over {\cF}}\right]\tPsi = 0
\label{weqn}
\end{equation}
Let us choose $\cW$ to be of the form $\cW = -i\tau + \cU(R,F)$. This ansatz is motivated by the fact
that $F'/2$ actually represents the radial mass-energy density within spherical shells. Inserting it
into (\ref{weqn}) gives the following equations for $\cU(R,F)$
\begin{eqnarray}
{{\partial \cU} \over {\partial R_*}} &=& \pm i {\sqrt{F \over {R-F}}},\cr
\cW &=& -i\tau \pm i \sqrt{F} \int dR {{\sqrt{R}} \over {R-F}},
\end{eqnarray}
outside the horizon ($R>F$), and
\begin{eqnarray}
{{\partial \cU} \over {\partial R_*}} &=& \pm i \sqrt{{F} \over {F-R}}\cr
\cW &=& -i\tau \pm i \sqrt{F} \int dR {{\sqrt{R}} \over {F-R}}
\end{eqnarray}
inside ($R<F$). In both regions the equations are easily integrated. They give
\begin{equation}
\cW = -i\tau \pm 2i \sqrt{F} \left[\sqrt{R} - \sqrt{F}\tanh^{-1}\sqrt{F\over R}\right]
\end{equation}
outside, and
\begin{equation}
\cW = -i\tau \mp 2i \sqrt{F} \left[\sqrt{R} - \sqrt{F}\tanh^{-1}\sqrt{R\over F}\right]
\end{equation}
inside.

Any collapse model that approximates reality must involve two or more regions of differing mass functions
(the simplest example would be a constant mass function representing the Schwarzschild exterior of a star
matched to another non-constant mass function representing the interior of the star). In general the mass
function must be continuous but it may not necessarily be differentiable across the boundary between these
regions. At such boundaries, the wave-functional must be appropriately matched, {\it i.e.,} required to be
both continuous and differentiable. Specific models which exhibit some of the interesting classical
features whose correct understanding cannot be had without appealing to quantum gravity, as mentioned
in the introduction, will be described in a forthcoming publication.

We have shown that there exists a canonical chart analogous to that used by Kucha\v r to describe the
Schwarzschild black hole and which describes the marginally bound collapse of inhomogeneous, pressureless
dust. This chart enjoys several advantages over the original. For one, the proper time of the collapsing
dust enters naturally and serves as a time variable in the Wheeler-DeWitt equation. Again, the chart
describes the collapse in terms of a more transparent and physically meaningful set of variables, yielding
a dynamical constraint that is greatly simplified and whose solution we have displayed over the entire
class of models being considered.

\section{Acknowledgements}

\noindent We acknowledge the partial support of FCT, Portugal, under contract number SAPIENS/32694/99.
L.W. was supported in part by the Department of Energy, USA, under contract Number DOE-FG02-84ER40153.


\end{document}